\documentclass[10ft,tightenlines,preprintnumbers,showpacs,
superscriptaddress,nofootinbib]{revtex4}
\usepackage{graphicx}
\usepackage{amssymb}
\usepackage{amsmath,mathrsfs,verbatim}
\usepackage{times}
\usepackage{latexsym}
\def\beq{\begin{equation}}
\def\eeq{\end{equation}}
\def\bey{\begin{eqnarray}}
\def\eey{\end{eqnarray}}

\def\lsim{\mathrel{\raise.3ex\hbox{$<$\kern-.75em\lower1ex\hbox{$\sim$}}}}
\def\gsim{\mathrel{\raise.3ex\hbox{$>$\kern-.75em\lower1ex\hbox{$\sim$}}}}

\newcommand{\be}{\begin{equation}}
\newcommand{\ee}{\end{equation}}

\newcommand{\mx}{m_\chi}
\newcommand{\ax}{\alpha_\chi}
\newcommand{\mphi}{m_\phi}

\begin{document}

\title{Self-interacting Dark Matter Benchmarks}

\author{Manoj Kaplinghat}
\affiliation{Department of Physics and Astronomy, University of California, Irvine, California 92697, USA}
\author{Sean Tulin}
\affiliation{Michigan Center for Theoretical Physics, University of Michigan, Ann Arbor, MI 48109, USA}
\author{Hai-Bo Yu}
\affiliation{Department of Physics and Astronomy, University of California, Riverside, California 92521, USA}

\date{\today}

\begin{abstract}
\vspace*{.0in}

Dark matter self-interactions have important implications for the distributions of dark matter in the Universe, from dwarf galaxies to galaxy clusters. We present benchmark models that illustrate characteristic features of dark matter that is self-interacting through a new light mediator.  These models have self-interactions large enough to change dark matter densities in the centers of galaxies in accord with observations, while remaining compatible with large-scale structure data and all astrophysical observations such as halo shapes and the Bullet Cluster.  These observations favor a mediator mass in the $1 - 100$ MeV range and large regions of this parameter space are accessible to direct detection experiments like LUX, SuperCDMS, and XENON1T.

\end{abstract}


\maketitle

Dark and visible matter have very different distributions in the Universe: dark matter (DM) forms diffuse halos (e.g., observed via graviational lensing maps), while visible matter undergoes dissipative dynamics and tends to clump into galaxies and stars.   However, this does not preclude the possibility of new dark sector interactions beyond the usual collisionless DM paradigm.  DM could have a large cross section for scattering with other DM particles and this scenario, dubbed self-interacting DM (SIDM)~\cite{1992ApJ...398...43C,Spergel:1999mh}, can affect the internal structure (mass profile and shape) of DM halos compared to collisionless DM.  In turn, astrophysical observations of structure, compared to numerical N-body simulations, can probe the self-interacting nature of DM.  It is worth emphasizing that tests of self-interactions can shed light on the nature of DM {\it even if DM is completely decoupled with respect to traditional DM searches}. We also note that the assumption in this note will be that the self-scattering is non-dissipative but it is possible for a sub-dominant fraction of dark matter to interact via dissipative processes \cite{Fan:2013tia}. 

There are, in fact, long-standing issues on small scales that may point toward SIDM.  Dwarf galaxies are natural DM laboratories since in these galaxies DM tends to dominate baryons well inside the optical radius. Observations indicate that the central regions of well-resolved dwarf galaxies exhibit cored profiles~\cite{Moore:1994yx,Flores:1994gz}, as opposed to steeper cusp profiles found in collisionless DM-only simulations~\cite{Navarro:1996gj}.  Cored profiles have been inferred in a variety of dwarf halos, including within the Milky Way (MW)~\cite{Walker:2011zu}, other nearby dwarfs~\cite{2011AJ....141..193O} and low surface brightness galaxies \cite{2008ApJ...676..920K}.  An additional problem concerns the number of massive dwarf spheroidals in the MW.  Collisionless DM simulations have a population of subhalos in MW-like halos that are too massive to host any of the known dwarf spheriodals but whose star formation should not have been suppressed by ultraviolet feedback~\cite{BoylanKolchin:2011dk}.  While these apparent anomalies are not yet conclusive -- e.g., baryonic feedback effects may be important~\cite{Governato:2012fa} -- recent state-of-the-art SIDM N-body simulations have shown that self-interactions can modify the properties of dwarf halos to be in accord with observations, without spoiling the success of collisionless DM on larger scales and being consistent with halo shape and Bullet Cluster bounds~\cite{Vogelsberger:2012ku,Rocha:2012jg,Peter:2012jh,Zavala:2012us}.  

The figure of merit for DM self-interactions is cross section per unit DM mass, $\sigma/m_\chi$, where $\chi$ is the DM particle.  To have an observable effect on DM halos over cosmological timescales, the required cross section per unit mass must be\footnote{Here, $\sigma$ refers to the momentum-transfer weighted cross section averaged over a Maxwellian velocity distribution for a given halo with characteristic (most probable) velocity $v_0$ .  See Ref.~\cite{Tulin:2013teo} for further details.}
\be
{\sigma}/{m_\chi} \sim 1 \; {\rm cm^2/g} \; \approx \; 2 \; {\rm barns/GeV} \, ,
\ee
or larger. 
From a particle physics perspective, this value is many orders of magnitude larger than the typical weak-scale cross section expected for a WIMP ($\sigma \! \sim\!1 \; {\rm picobarn}$).  Evidence for self-interactions would therefore point toward a new dark mediator particle $\phi$ that is much lighter than the weak scale.  Such light mediators have been invoked within a variety of other DM contexts as well, including explaining various indirect detection anomalies; see e.g.~\cite{Feng:2008ya,Pospelov:2008jd, ArkaniHamed:2008qn}.

As one example, DM self-interactions can arise if DM is coupled to a massive dark photon $\phi$ from a hidden $U(1)^\prime$ gauge symmetry~\cite{Feng:2009mn,Ackerman:2008gi,Feng:2009hw,Buckley:2009in,Loeb:2010gj,Tulin:2012wi,Tulin:2013teo}.  Other examples where dark matter self-interactions arise include mirror dark matter \cite{Mohapatra:2001sx,Foot:2004wz,Foot:2012ai} and atomic dark matter \cite{Kaplan:2011yj,CyrRacine:2012fz}, both appearing in the framework of hidden sector dark matter. The non-relativistic self-scattering mediated by a dark photon can be described by a Yukawa potential, 
\begin{eqnarray} \label{yukawapot}
V(r)=\pm\frac{\ax}{r} \, e^{-\mphi r}, 
\end{eqnarray} 
where $\alpha_\chi$ is the ``dark fine structure constant.''  For symmetric DM (both $\chi,\bar \chi$ are present today) scattering can be repulsive ($+$) or attractive ($-$), while for asymmetric DM (only $\chi$ is present today) scattering is purely repulsive.  Given the potential in Eq.~\eqref{yukawapot}, the cross section $\sigma$ can be computed using standard methods from quantum mechanics as a function of the three parameters $(m_\chi, m_\phi, \alpha_\chi)$ and the relative velocity $v$~\cite{Tulin:2013teo}.  

Different size DM halos have different characteristic velocities, giving complementary information about $\sigma(v)$.  
Similar to Rutherford scattering, DM self-scattering through a light mediator is typically suppressed at large velocities compared to smaller velocities.  Therefore, it is natural for DM to be self-interacting in dwarf halos, while appearing to be collisionless in larger halos.  For example, the Bullet Cluster is often quoted as an example of an observation that categorically rules out self-interactions in the dark sector.  This is not true since the relative velocity in the Bullet Cluster system ($v \approx 3000\; {\rm km/s}$) is much larger than in dwarf halos ($30 \; {\rm km/s}$).  As we show below, this constraint, while important, eliminates only a small region of SIDM parameter space.

Aside from self-interactions, the mediator $\phi$ can also set the DM relic density in the early Universe through $\chi\bar{\chi}\rightarrow\phi\phi$ annihilation.  For symmetric DM, 
the required annihilation cross section is $\left<\sigma v\right>_{\rm ann} \approx 5\times10^{-26}~{\rm cm^3/s}$, which fixes 
$\ax\approx 4\times10^{-5}(\mx/{\rm GeV})$.   For asymmetric DM, although the relic density is determined by a primordial asymmetry, $\langle \sigma v \rangle_{\rm ann}$ has to be {\it larger} than in the symmetric case, implying $\ax\gtrsim 4\times10^{-5}(\mx/{\rm GeV})$.

\begin{figure}
\includegraphics[scale=0.63]{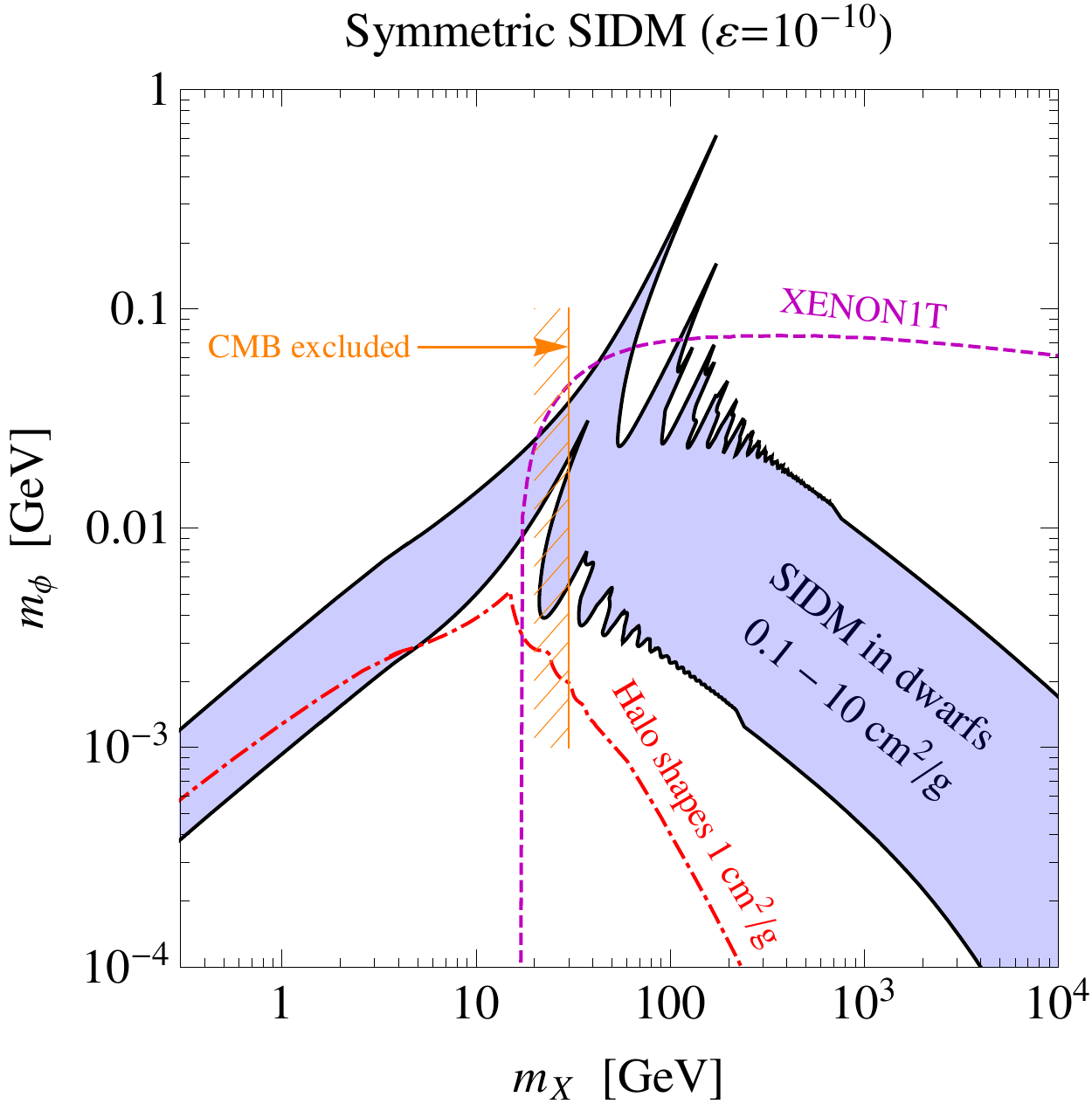}
\includegraphics[scale=0.63]{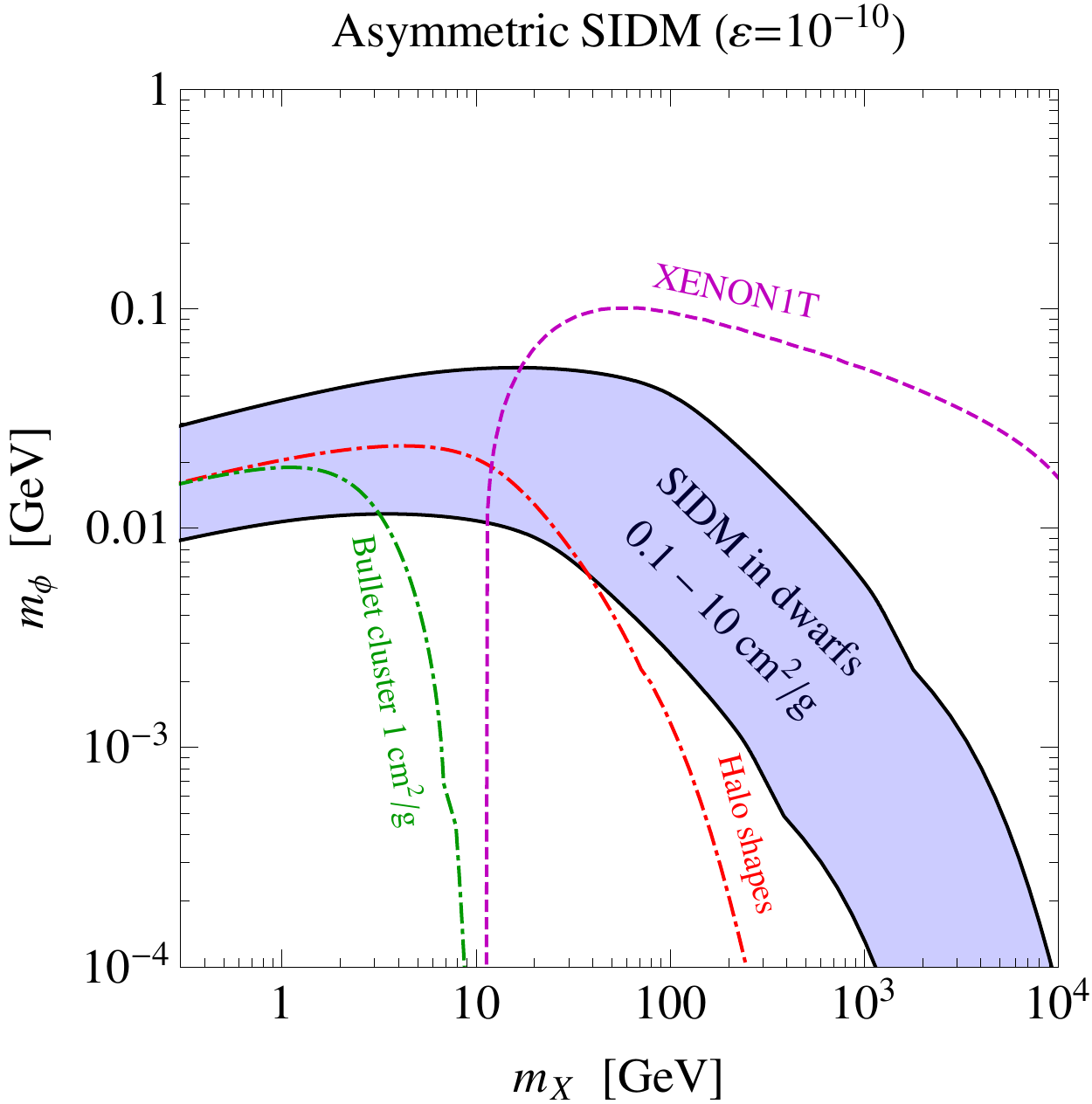}
\caption{Parameter space for SIDM $\chi$ with a vector mediator $\phi$, as a function of their masses $m_\chi,m_\phi$, for symmetric DM with $\alpha_\chi$ fixed by relic density (left) and asymmetric DM with $\alpha_\chi = 10^{-2}$ (right). Shaded region indicates the region where DM self-interactions would lower densities in the central parts of dwarf scales consistent with observations. The upper (lower) boundary corresponds to $\langle \sigma_T\rangle/m_\chi=0.1~{\rm cm}^2/{\rm g}$ ($10~{\rm cm}^2/{\rm g}$). Dot-dashed curves show halo shape constraints on group scales ($\sigma/m_\chi < 1 \; {\rm cm^2/g}$) and the Bullet Cluster constraint ($\sigma/m_\chi < 1 \; {\rm cm^2/g}$).  Dashed lines show direct detection sensitivity for XENON1T if $\phi$ has kinetic mixing with the photon with $\epsilon=10^{-10}$. The vertical hatched boundary shows exclusion from CMB if $\phi \to e^+ e^-$. See text for details. }
\label{paramspace}
\end{figure}

Fig.~\ref{paramspace} shows the parameter space for this SIDM model as a function of $m_\chi$ and $m_\phi$.  The left panel corresponds to symmetric DM, where $\alpha_\chi$ is fixed by relic density, while the right panel corresponds to asymmetric DM with $\alpha_\chi = 10^{-2}$.  The shaded regions show where SIDM can explain halo anomalies on dwarf scales, with a generous range of cross section $0.1 \lesssim \sigma/m_\chi \lesssim 10\; {\rm cm^2/s}$ and taking a characteristic velocity $v_0 = 30 \; {\rm km/s}$. The upper (lower) boundary corresponds to $\langle \sigma_T\rangle/m_\chi=0.1~{\rm cm}^2/{\rm g}$ ($10~{\rm cm}^2/{\rm g}$). 
To implement the Bullet Cluster constraint, we require $\sigma/m_\chi \lesssim 1 \; {\rm cm^2/g}$ for a relative velocity $v \approx 3000\; {\rm km/s}$~\cite{Randall:2007ph}, shown by the green dot-dashed contour.  Other constraints arise from the ellipticity of DM halos of groups of galaxies; we require $\sigma/m_\chi \lesssim 1 \; {\rm cm^2/s}$ for halos of characteristic velocity $v_0 \approx 300 \; {\rm km/s}$~\cite{Peter:2012jh}, shown by the red dot-dashed contour (``Halo shapes'').  From these bounds, the low ($m_\chi,m_\phi$) region is excluded in Fig.~\ref{paramspace}.

The dark and visible sectors need not be completely decoupled.  For example, if there exist new states charged under both the Standard Model (SM) and $U(1)^\prime$ gauge symmetries, mixing can arise between $\phi$ and the photon or $Z$ boson.  This generates effective couplings of $\phi$ to protons and neutrons, giving rise to signals in direct detection experiments.  
In the limit of zero momentum transfer, the spin-independent (SI) $\chi$-nucleon cross section can be written as
\be
\sigma_{\chi n}^{\rm SI} = \frac{16\pi \ax \alpha_{\rm em} \epsilon_{\rm eff}^2 \mu^2_{\chi n}}{m^4_\phi}  \approx 10^{-24} \; {\rm cm}^2 \times \epsilon_{\rm eff}^2 \left(\frac{30 \; {\rm MeV}}{m_\phi} \right)^4 \times \left\{ \begin{array}{cc} ({m_\chi}/{200 \; {\rm GeV}}) & {\rm symmetric \; DM} \\
({\alpha_\chi}/{10^{-2}}) & {\rm asymmetric \; DM} \end{array} \right. ,
\label{eq:dd}
\ee
where $\mu_{\chi n}$ is the $\chi$-nucleon reduced mass, $\alpha_{\rm em}$ is the electromagnetic fine structure constant, and $\epsilon_{\rm eff}$ is the effective $\phi$-nucleon coupling, normalized to the proton electric charge $e$. Since SIDM prefers a very light mediator, with mass $m_\phi \sim 1 - 100$ MeV, it is clear that direct detection experiments are sensitive to very small couplings $\epsilon_{\rm eff}$. 

As an example, we consider the case of kinetic mixing between $\phi$ and the photon, governed by the parameter $\epsilon$~\cite{Holdom:1985ag}.  This mixing induces a coupling of $\phi$ to SM particles carrying electric charge, so that $\phi$ decays predominantly to $e^+ e^-$ for $m_\phi$ in the $1 - 100$ MeV range prefered for SIDM.  The direct detection cross section is governed by the $\phi$-proton coupling with $\epsilon_{\rm eff} = \epsilon Z/A$, where $Z/A$ is the proton fraction of the target nucleus.  However, there are various constraints on the $\epsilon$.  Late decays of $\phi$ can inject energy to the plasma and modify standard big bang nucleosynthesis in the early Universe. 
Requiring the $\phi$ lifetime to be longer than $\sim 1$ second for leptonic decay modes, we derive a lower bound $\epsilon\gtrsim10^{-10}\sqrt{10~{\rm MeV}/\mphi}$~\cite{Lin:2011gj}. 
The upper bound from the low energy beam dump experiments is $\epsilon\lesssim10^{-7}$ for $\mphi\lesssim400~{\rm MeV}$~\cite{Bjorken:2009mm}, while the region $10^{-10} \lesssim \epsilon \lesssim 10^{-7}$ is excluded for $\mphi\lesssim100~{\rm MeV}$ by energy loss arguments in supernovae~\cite{Dent:2012mx} (although this constraint depends sensitively on assumptions about the temperature and size of the supernova core).  Regardless, for what follows, we take $\epsilon = 10^{-10}$ as a benchmark point. 

Since the mediator mass $\mphi\sim1-100$ MeV is comparable or less than the typical momentum transfer $q\sim50$ MeV in nuclear recoils, nuclear recoil interactions for SIDM are momentum-dependent and cannot be approximated by a contact interaction~\cite{Feldstein:2009tr,Chang:2009yt}. Here, we take a simplified approach by multiplying the total $q^2=0$ DM-nucleus cross section by a $q^2-$ dependent form factor: $\sigma^{\rm SI}_{\chi N}(q^2)=\sigma^{\rm SI}_{\chi N}(q^2=0)f(q^2)$, with $f(q^2)=\mphi^4/(\mphi^2+q^2)^2$. We take a fixed value $q=50$ MeV for Xenon and assume that the cross section limits quoted in the XENON experiment apply to $\sigma^{\rm SI}_{\chi N}(q^2)$ directly. We have checked that our simple approximation can reproduce the XENON100 reanalysis in~\cite{Fornengo:2011sz}. In Fig.~\ref{paramspace}, we show how direct detection sensitivities from XENON1T~\cite{Aprile:2012zx} map onto SIDM parameter space for $\epsilon = 10^{-10}$ and $Z/A \approx 0.4$ (purple dashed contours). It is interesting to note that the current XENON100~\cite{Aprile:2012nq} limits are not sensitive to SIDM with $\epsilon=10^{-10}$ because of the suppression from $f(q^2)$.

For symmetric DM, residual annihilation can lead to additional reionization around the recombination epoch via $\chi \bar \chi \to \phi \phi \to e^+ e^- e^+ e^-$, which is constrained by CMB observations~\cite{Galli:2009zc,Slatyer:2009yq}. For ${\rm BR}(\phi \to e^+ e^-)=1$, symmetric SIDM is excluded for $\mx$ below $\sim30~{\rm GeV}$~\cite{Lopez-Honorez:2013cua}, as indicated in Fig.~\ref{paramspace} (left) with the vertical hatched boundary. For asymmetric DM, this constraint does not apply.  (We also note that this bound is weakened if $\phi$ decays to neutrinos, which occurs if $\phi$ mixes with the $Z$ boson.)  There are further connections to indirect searches (such as AMS-02 and Fermi) and these aspects will be discussed in a paper in preparation \cite{KLY} in the context of models with kinetic mixing.


\begin{figure}
\includegraphics[scale=0.63]{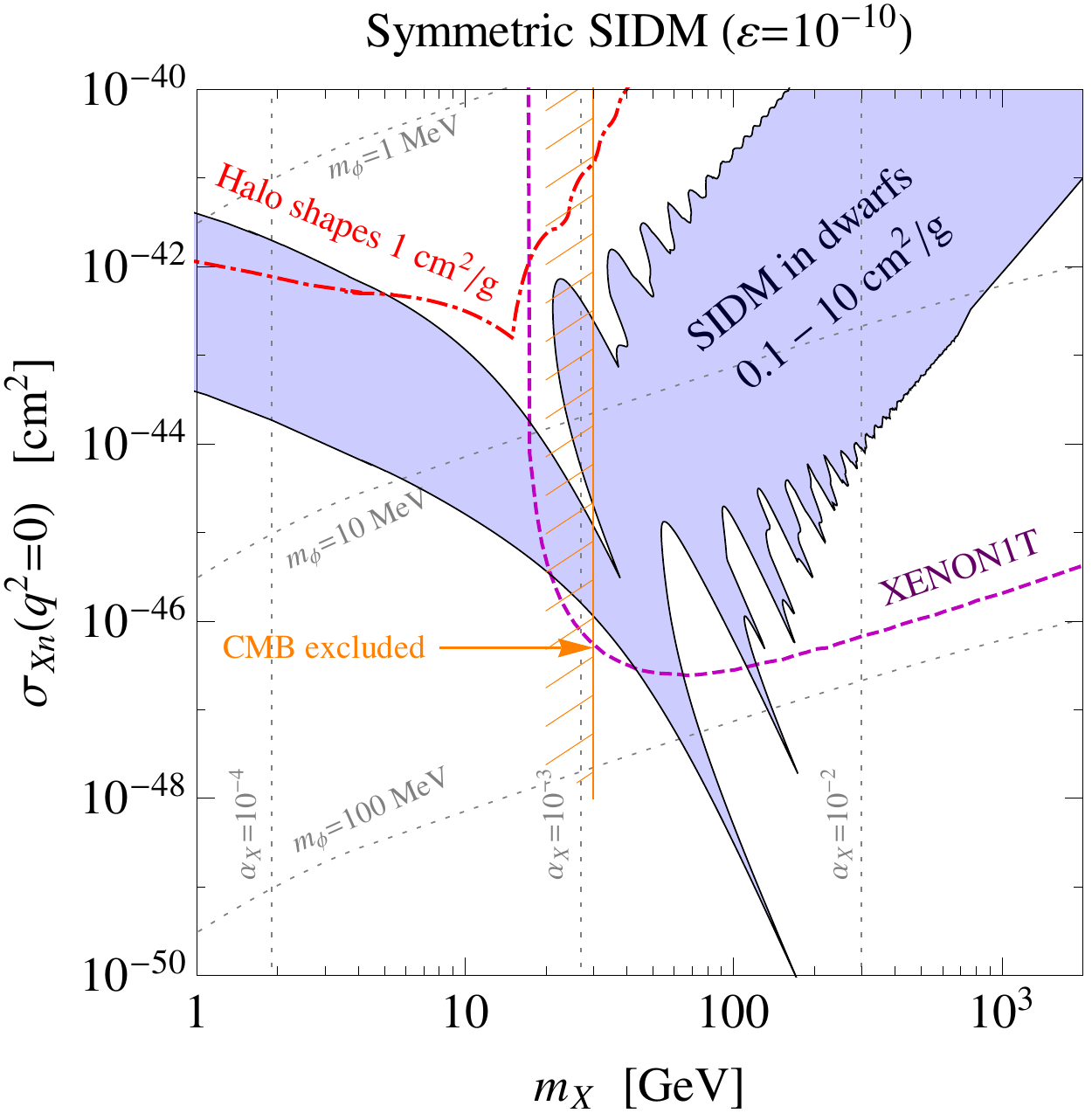} \includegraphics[scale=0.63]{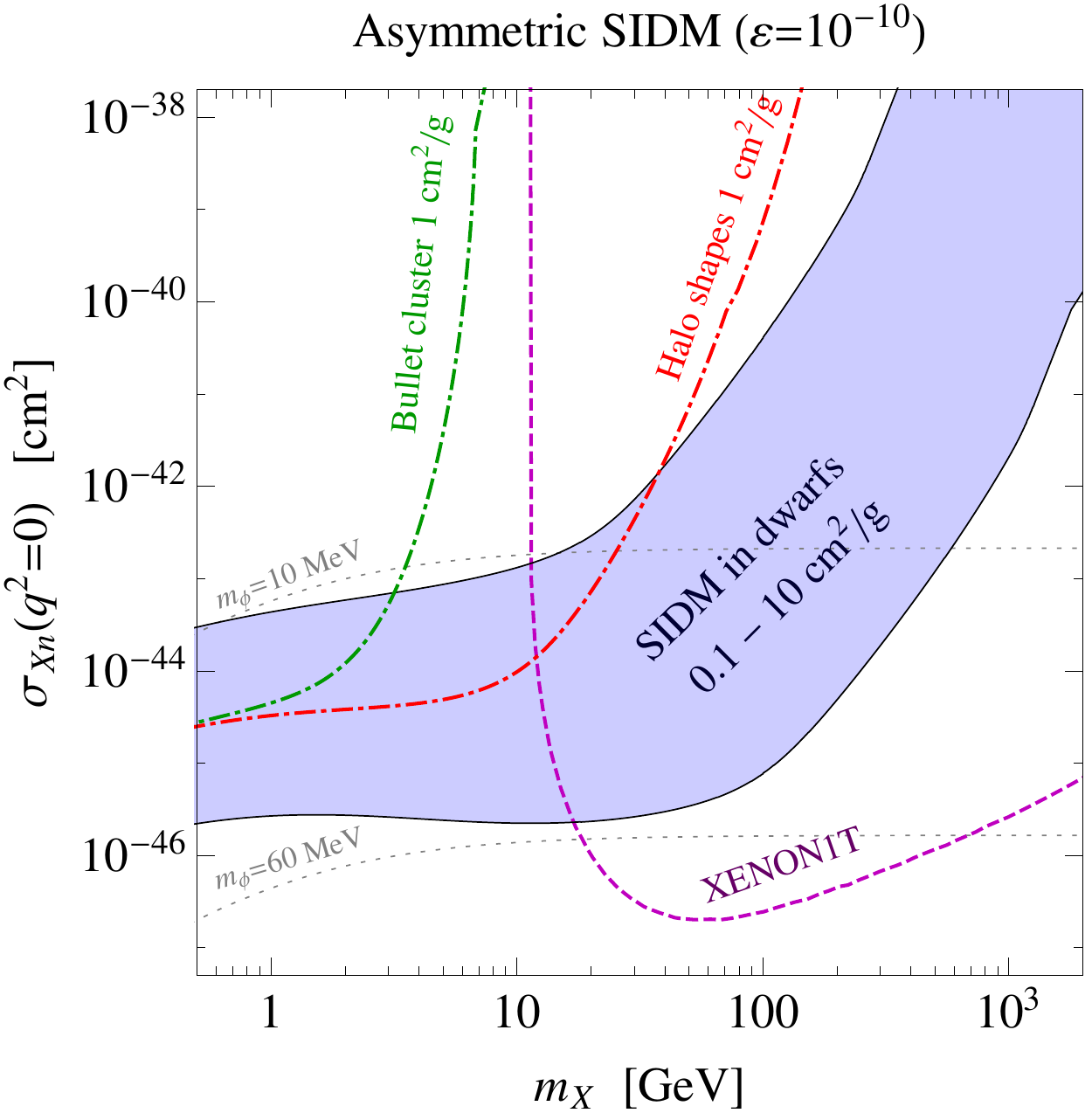}
\caption{Prospects for direct detection of self-interacting DM that couples to quarks via gauge kinetic mixing with $\epsilon=10^{-10}$.  Left figure is for symmetric DM, with $\alpha_X$ fixed by relic density constraints; right figure is asymmetric DM, with $\alpha_X = 10^{-2}$.  Shaded region indicates the region where DM self-interactions would lower densities in the central parts of dwarf scales consistent with observations. The upper (lower) boundary corresponds to $\langle \sigma_T\rangle/m_\chi=10~{\rm cm}^2/{\rm g}$ ($0.1~{\rm cm}^2/{\rm g}$). Direct detection sensitivity from future XENON1T experiments shown by dashed curves. Astrophysical limits from halo shapes and the Bullet Cluster shown by dot-dashed lines.  The range of $m_\phi, \alpha_X$ values are shown by dotted lines. The vertical hatched boundary shows exclusion from CMB if $\phi\rightarrow e^+e^-$ (left). These figures are taken from a work in preparation by Ref.~\cite{KTY} where the complementarity in models with kinetic mixing and other ways of connecting to the SM are explored.}
\label{directdet}
\end{figure}

In Fig.~\ref{directdet}, we illustrate the complementarity between astrophysical probes and direct detection in constraining SIDM.  Fixing $\epsilon = 10^{-10}$, we show the SIDM prediction for SI scattering cross section per nucleon in direct detection experiments for both symmetric DM (left) and asymmetric DM (right).  As in Fig.~\ref{paramspace}, the shaded band shows the preferred parameter region for solving dwarf-scale anomalies, while the red and green contours denote limits from halo shape observations and the Bullet Cluster, respectively. The purple dashed lines show the projected XENON1T bounds~\cite{Aprile:2012zx}. The dotted gray lines denote contours of constant $\alpha_\chi$ and $m_\phi$.  

In summary, SIDM is a well-motivated DM scenario and it generically predicts a 1-100 MeV dark force carrier. When it couples to the SM sector, it generates signals in direct and indirect detection experiments. In the benchmark models we consider, where the dark sector couples to the SM through kinetic mixing with parameter $\epsilon=10^{-10}$, current direct detection experiments are not sensitive to SIDM. But, future direct detection experiments, such as LUX~\cite{Akerib:2012ys}, SuperCDMS~\cite{Brink:2012zza}, and XENON1T will offer great sensitivity to detect SIDM, with XENON1T experiment covering most of the parameter space for SIDM masses greater than about 20 GeV.  The simple but generic example considered in this note demonstrates that astrophysical observations and direct detection experiments complement each other in the search for SIDM candidates. 

{\it Acknowledgments}: We thank Rafael Lang for useful discussions. MK is supported by NSF Grant No.~PHY-1214648. ST is supported by the DOE under contract de-sc0007859. HBY is supported by startup funds from the UCR.

\bibliography{snowbib}

\end{document}